# Relaxation of inverted quantum system coupled with metallic nanoobjects

## A. Chipouline

**Abstract.**

The problems of the luminescent enhancement and life time shortening for the inverted quantum system (QS) are analysed. The commonly accepted in publications approach appears to be questionable from the point of view of basic principles. An alternative approach using the well-known density matrix formalism and kinetic equation for the electrons in the coupled nano object is suggested.

## 1. Introduction

Here I consider the problem of the relaxation dynamics of a two level quantum system (QS) in the frame of the developed in [Chipouline 12 JOPT] approach and compare it with the known and commonly accepted math tools used to describe this problem. I show that the consideration of the qualitative picture of the relaxation processes leads to particular concerns about the validity of the commonly accepted approaches.

The relaxation time modification has been considered in many different realizations (including photonic crystals [Maragkou 12] and plasmonic waveguides [Barthes 11]), but here to be precise we consider only the coupling between the localized plasmons and two-level QSs.

The problem of the life time modification of the QS due to the presence of plasmonic nanostructures has a long history. Since seminal paper of Purcell [Purcell 46], and (in more than 30 years) fundamental theoretical works [Ruppin 82], [Gersten 81], several excellent reviews [Klimov 01], [Giannini 11], [Dregely 12] and books about Cavity Quantum Electrodynamics [Berman 94], [Haroche 92] have been published; a full and comprehensive publication reference list can be found there. Among the recent achievements it should be mentioned the experimental realization of the plasmonic patch antenna for control of the spontaneous emission [Belacel 13] (which has been theoretically considered in [Esteban 10]) and observation of the ultrafast Rabi oscillation [Vasa 13], which manifestoes the strong coupling regime [Vlack 12], [Trügler 08]. The strong coupling assumes positioning of the quantum system in close proximity to the nanoresonator, which in turns requires inclusion of higher (than dipole) order modes in nanoresonators and dipole-forbidden transitions in the QSs [Kern 12_1], [Stobbe 12], [Filter 12]. Coupling with realistic shapes of the nanoresonators has been considered in [Kern 11] and [Kern 12_2]. The peculiarities of the disordered plasmonic on the radiative and nonradiative relaxation rates have been considered in [Caze 12], and influence of the plasmonic cloak has been investigated in [Kort 13]. Accurate calculation of the Green function for complex plasmonic structures allowed estimating the spatially resolved Purcell effect for hyperbolic metamaterials [Poddubny 12] and wire metamaterials [Poddubny 13]. The relaxation time modification due to the presence of the nanoresonator [Stockman 10], has been taken into account in the modelling of the spaser dynamics [Bergman 03], [Stockman 11] using basically the same approach as in [Novotny 06].

## 2. The accepted approach to the estimation of Purcell effect

Analysis of the "state of the art" of the activity in area of the nanoplasmonic assisted relaxation time modification would worth to start from several definitions, namely – what exactly we name "Purcell effect". From the experimental point of view, the observed in the tests relaxation time modification is called Purcell effect without any details about an actual functional form of the relaxation curves and discussion about relaxation channels. In most cases (not only in experimental, but also in vast majority of the theoretical publications) the Purcell effect itself is not distinguished with the Purcell factor, which is the ratio between the relaxation rates in vacuum and near the plasmonic nanostructures. This assumes that the relaxation follows exponential time dependence in both cases, namely in free space and in the case of coupling with the nanoresonators. To my opinion, there is neither theoretical nor experimental evidence that the relaxation in general has an exponential shape; moreover, both experimental [Giannini 11], [Belacel 13] and theoretical results [Mertens 07] show the non-exponential decay law. In addition, it has to be mentioned that the evaluation expressions for the Purcell coefficient, based on the estimated values for the effective mode volume and resonator quality factor, in contrast to the microcavities does not give correct estimation for the plasmonic nanoantenna [Koenderink 10]. Hereafter, I do not apriori assume that the decay follows an exponential law and show that the exponent is a good approximation only in the case of relaxation caused by interaction with stochastic environment.

The relaxation dynamics of the QS interacting with a plasmonic nanoresonator can be considered classically and quantum mechanically, we start from the latter one. Due to methodological reasons I would like to remind basics of the relaxation in a free space. First, it has to be realized that in the frame of Schrödinger equation for the QS in free space there is no relaxation until there is no external fields. The relaxation appears only as a result of consideration of the vacuum fluctuations of the electromagnetic field (second quantization). Here it is worth to remind that the spontaneous emission cannot be considered as just influence of the vacuum fluctuations on the QS in the frame of the first perturbation order, but rather as a coupled dynamics of the system "field plus QS". This fundamental question has been perfectly explained in [Ginzburg 83], see also different points of view on this problem in [Weisskoppf 35], [Weisskpopf 81] and [Fermi 32]. The relaxation process itself remains classic effect (in analogy with, for example energy relaxation of an accelerated charge), while the vacuum fluctuations is of course a purely quantum one.

A derivation of the spontaneous decay rate in the frame of the quantum electro-dynamic approach can be found in textbooks, for example in [Novotny 06]. Consideration of the coupled "electric field plus QS" transition rate $\gamma$ according to the Fermi's Golden Rule gives finally:

$$\gamma = \frac{2\omega}{3\hbar\varepsilon_0}|\mu|^2 \rho_\mu(\vec{r}_0, \omega_0)$$

$$\rho_\mu(\vec{r}_0, \omega_0) = 3\sum_k \left[\vec{n}_\mu \cdot (\vec{u}_k \vec{u}_k^*) \cdot \vec{n}_\mu\right] \delta(\omega_k - \omega_0) \tag{1}$$

Here $\vec{\mu}$ is the dipole moment of the transition of the QS, $\vec{\mu} = \mu \vec{n}_\mu$, $\vec{r}_0$ is position of the QS, $\rho_\mu(\vec{r}_0, \omega_0)$ is the partial local density of states (LDOS), $\vec{u}_k$ are the normal modes, $\vec{n}_\mu$ is the unit vector in the direction of $\vec{\mu}$. The electric field is expanded over spatially dependent positive and negative frequency parts:

$$\vec{E} = \sum_k \left[ \vec{E}_k^+ a_k(t) + \vec{E}_k^- a_k^\dagger(t) \right]$$
$$\vec{E}_k^+ = \sqrt{\frac{\hbar \omega_k}{2\varepsilon_0}} \vec{u}_k, \quad \vec{E}_k^- = \sqrt{\frac{\hbar \omega_k}{2\varepsilon_0}} \vec{u}_k^*$$
(2)

where:

$$a_k(t) = a_k(0) \exp(-i\omega_k t), \quad a_k^\dagger(t) = a_k^\dagger(0) \exp(i\omega_k t)$$
(3)

The relaxation dynamics of the probability of the QS to be found in the excited state (population of the excited state) decays exponentially. Decay rate (1) describes relaxation into the vacuum field modes $\vec{u}_k$. It is important to underline at this point two facts. First, this decay rate assumes the radiation rate, which means that all the transitions give one photon per transition into the far or near field zones depends on what kind of modes $\vec{u}_k$ are: if the QS is situated in the near/far field zone, then $\vec{u}_k$ are the near/far field zone modes. Second, the interaction with vacuum field has a stochastic character: the vacuum fluctuations occur randomly with some probability, and quantum character of these fluctuations is in the fact, that they can absorb a photon (transition from the state with zero photons to the state with one photon), but cannot give photon (probability of the reverse transition, from the state with zero photons to the state with one photon is zero). Due to this fact there is no "pumping" process of the QS due to the absorption of the vacuum photons. This important quantum property leads to the appearance of the relaxation in the first order of perturbation expansion – if the fluctuation would have a classical nature, the relaxation would appear in the second order of perturbation. We see later, that this differentiates the relaxation due to the vacuum fluctuations and due to the interaction with the environmental classical fluctuations.

The problem now is reduced to the finding of the functional form for the LDOS. There is an elegant way using the methods of Green functions [Novotny 06], which the vast majority of the recently appeared papers are based on. Remind that at the elaboration of (1) the orthogonality of the ground states was used (see [Novotny 06] for details), but the orthogonality of the filed eigen modes $\vec{u}_k$ was not assumed. The eigen functions $\vec{u}_k$ satisfy Helmholz equation, which is for generally inhomogeneous and dispersive (and consequently due to the causality principle lossy) media is:

$$\nabla \times \nabla \times \vec{u}_k(\vec{r}, \omega_k) - \varepsilon(\vec{r}, \omega_k) \frac{\omega_k^2}{c^2} \vec{u}_k(\vec{r}, \omega_k) = 0$$
$$\nabla \times \nabla \times \vec{G}(\vec{r}_0, \vec{r}, \omega_k) - \varepsilon(\vec{r}, \omega_k) \frac{\omega_k^2}{c^2} \vec{G}(\vec{r}_0, \vec{r}, \omega_k) = \vec{I} \delta(\vec{r}_0 - \vec{r})$$
(4)

Here Green function $G(\vec{r}_0, \vec{r}, \omega_k)$ is introduced and $\bar{I}$ is the unit dyad. At the elaboration of the expression for the LDOS through the Green function, it is necessary to assume that the modes $\vec{u}_k$ are orthogonal, which is straightforward for any media without losses, i.e. constant and real permittivity $\text{Re}\{\varepsilon(\vec{r}, \omega_k)\} \neq 0, \text{Im}\{\varepsilon\} = 0$. In this case the $\vec{u}_k$ are orthogonal:

$$\int \vec{u}_k(\vec{r}, \omega_k) \vec{u}_{k'}^*(\vec{r}, \omega_{k'}) d^3\vec{r} = \delta_{kk'} \tag{5}$$

and expressions for the relaxation rate and LDOS are:

$$\gamma = \frac{2\omega_0}{3\hbar\varepsilon_0} |\mu|^2 \rho_\mu(\vec{r}_0, \omega_0)$$

$$\rho_\mu(\vec{r}_0, \omega_0) = \frac{6\omega_0}{3\hbar\varepsilon_0} \left[ \vec{n}_\mu \cdot \text{Im}\{\vec{G}(\vec{r}_0, \vec{r}_0, \omega_0)\} \cdot \vec{n}_\mu \right] \tag{6}$$

Finally, LDOS for the homogeneous isotropic media $\text{Re}\{\varepsilon(\vec{r}, \omega_k)\} = \varepsilon(\omega_k), \text{Im}\{\varepsilon\} = 0$ takes well-known form of the black-body radiation:

$$\rho_\mu(\omega_0) = \varepsilon \frac{\omega_0^2}{\pi^2 c^3} \tag{7}$$

And the decay rate is:

$$\gamma = \frac{\varepsilon^{3/2} \omega_0^3 |\mu|^2}{3\pi\varepsilon_0 \hbar c^3} \tag{8}$$

In the case of a dielectric inhomogeneous media ($\varepsilon(\vec{r}, \omega), \text{Im}(\varepsilon(\vec{r}, \omega)) = 0$ which is a good approximation for dielectric in transmission windows) the procedure remains the same excepting more complicated Green function and consequently spatially dependent LDOS and decay rate.

In the case of significant losses (as it takes place for the metallic nanoresonators) orthogonality condition (5) fails due to the nonzero imaginary part of the permittivity, which makes (4) nonhermitian. More sophisticated math tools have been suggested for this case [Dung 00], [Knoll 00]; moreover, applicability of the electrodynamic approach was shown to be valid for the case of the very small distances from the metallic surface [Yeung 96]. Nevertheless, wide application of the Green function in its original formulation (6) seems to be stipulated at the moment rather by its relative simplicity than by a full justification of this method for the metallic (dispersive and lossy) nanoresonators [Poddubny 12], [Poddubny 13]. In the frame of the same electrodynamic approach it is rather straightforward to calculate radiative and nonradiative decay rates. Both classical [Ruppin 82] and electrodynamic approaches [Gersten 81] have been proposed. Taking into account radiative losses in addition to the Joule ones for the nanoresonators, the radiative decay rate can be calculated by energy flux through the surface of the sphere surrounded both the QS and nanoresonator, while the nonradiative decay rate is determined by the calculating of the power dissipated in the nanoresonator due to the Joule heating [Mertens 07].

## 3. Concerns about the commonly accepted approach

From the other side, the physical picture behind this approach appears to be to some extend controversial.

First, this picture being applied for the near field (zero wave vectors) contradicts to the relativistic principles, which is not surprising and most probably has to be accepted.

Second, all math manipulations with the Green functions and eigen modes assumes that the conditions for the eigen modes to be excited are fulfilled, and that all free electrons in the nanoparticle participate in this mode move coherently. This assumption does not seem to be realistic. In fact, the plasmonic mode relaxation time is about 100 fs, which is comparable with the coherence time of the spontaneous photons, and under this condition the eigen mode just does not have enough time to be formed. This ambiguity is connected with the logic jump between the expansion of the filed over eigen functions (4) for the free homogeneous and inhomogeneous spaces. In fact, for the homogeneous space the microscopic Maxwell equations are used, while introduction of permittivity assumes that we work with macroscopic Maxwell equations. At this transition (homogenization procedure) the electron dynamics is assumed to be coherent, which is questionable for the considered case of a single photon.

Another concern about the use of the eigen mode appears when we consider an elementary act of the absorption. In fact, one particular photon gets absorbed just by one particular electron, which is a free one, because of the dynamics of the electron does not depend on the nano sizes of the nanoresonator – dynamics of the electron is not quantized and does not influence dynamics of the other electrons (model of free electron gas without interaction).

One more concern is connected with the discussed above peculiarities of the vacuum fluctuations, which due to the quantum nature does not possess process of absorption of the vacuum fluctuation photons by a QS, which leads to the appearance of the relaxation effect in the first order of the perturbation series. The self-influence, described by the Green function at its origin (see [Novotny 06]), is already the second order of the perturbation theory which evidently exceeds accuracy of the model (remind that the relaxation process is described in the frame of the first perturbation order). It is worth noting here that in the frame of the density matrix approach (will be considered later) the relaxation processes due to the interaction with a thermo bath (classical fields) is described by the second order perturbation terms. In the frame of the Green function approach, the first order perturbation would correspond to the influence of the nanoresonator on the LDOS by taking the Green function not at the origin of the QS, but rather between the nanoresonator and the QS $\vec{G}(\vec{r}_0,\vec{r}_0,\omega_k) \to \vec{G}(\vec{r}_{NR},\vec{r}_0,\omega_k)$, here $\vec{r}_{NP}$ is the center of the nanoresonator. This qualitative discussion shows that the interpretation of the relaxation rate modification in terms of the classical notations most probably requires further corrections.

## 4. Can a quantum dynamics be described by a harmonic oscillator equation?

Now, let us briefly consider the classical approach. In the frame of this approach the quantum dynamics is described by a harmonic oscillator equation. Let us first investigate under which approximation this approach remains valid. We start from the density matrix equation with relaxation in the basis of the eigen function of the unperturbed Hamiltonian for two level system with pump (see [Chipouline 12 JOPT] for details), namely:

$$\begin{cases} \dfrac{d\rho_{12}}{dt} + i\omega_{12}\rho_{12} + \dfrac{\rho_{12}}{\tau_2} = -\dfrac{iH_{12}N}{\hbar} \\ \dfrac{dN}{dt} + \dfrac{N - N_0}{\tau_1} = -\dfrac{2iH_{12}(\rho_{12} - \rho_{12}^*)}{\hbar} \\ \tau_1 = \dfrac{\tilde{\tau}_1}{W\tilde{\tau}_1 + 1} \end{cases} \quad (9)$$

Here $N = \rho_{22} - \rho_{11}$ and $N_0 = \dfrac{(W\tilde{\tau}_1 - 1)}{(W\tilde{\tau}_1 + 1)}$, $\rho_{22}$, $\rho_{11}$ and $\rho_{12}$, $\rho_{12}^*$ are the diagonal and non-diagonal matrix density elements, respectively; $\tau_2$ and $\tilde{\tau}_1$ are the constants describing phase and energy relaxation processes due to the interaction with a thermostat; $\omega_{21} = (E_2 - E_1)/\hbar$ is the transition frequency between levels 2 and 1; $H_{12}$ is the Hamiltonian matrix element responsible for interaction of QS with the external fields; $W$ is the phenomenological pump rate – this could model pumping QS.

Introducing $P = \rho_{12} + \rho_{12}^*$, one can finally reduce (9) to the following form:

$$\begin{cases} \dfrac{d^2 P}{dt^2} + \dfrac{2}{\tau_1}\dfrac{dP}{dt} + \left(\dfrac{1}{\tau_1^2} + \omega_{21}^2\right)P = \dfrac{2\omega_{21}H_{12}N}{\hbar} \\ \dfrac{dN}{dt} + \dfrac{N - N_0}{\tau_2} = -\dfrac{2H_{12}}{\hbar\omega_{21}}\left(\dfrac{dP}{dt} + \dfrac{P}{\tau_1}\right) \end{cases} \quad (10)$$

The first equation in (10) could be taken as a trivial harmonic oscillator one in case when these two equations are separable, i.e. $N$ does not depend on $P$. It could be a good approximation for the unpumped QS without saturation, i.e. under low intensity external Hamiltonian $H_{12}$; in this case $N \approx N_0 = -1$. In contrast, in case of the relaxation dynamics it is assumed that $N$ evolves from $N = 1$ to $N = -1$, which excludes any reasonable justification for the separation of the first equation as a harmonic oscillator one.

In case of a harmonic oscillator equation the damping parameter is supposed to be responsible for the both line bandwidth and the relaxation time. Let us consider a typical quantum dot with the typical bandwidth of about 50 nm around central wavelength of 1000 nm. Inverse bandwidth spectrum width gives us the approximated relaxation time:

$$\tau \approx \dfrac{1}{\Delta\nu} = \dfrac{\lambda^2}{c\Delta\lambda} \sim 100\,fs \quad (11)$$

From the other side, the measured in the tests relaxation time is typically about 1-10 ns. In the frame of the density matrix approach there is no any contradictions, because of there are two different times for the spectrum width (phase relaxation time) and for the energy relaxation time, namely $\tau_2$ and $\tau_1$. The relaxation time, which can be calculated using (1) is the energy relaxation time $\tau_1$, because of this time assumes energy transition with a photon emission; the $\tau_2$ describes change in the phase of the eigen state without energy transition. Clear, that the harmonic oscillator equation neither can be elaborated within some reasonable approximations, nor can give even qualitative explanation for the different relaxation times and spectrum bandwidth of the QS. It has to be pointed out, that this is again consequence of the quantum character of the QS. For example, in case of the relaxation of the plasmonic oscillations the relaxation time (again about 100 fs) matches pretty good with the observed bandwidth of the plasmonic resonances which is again comparable with one for the QDs and is typically about 30-50 nm.

## 5. Relaxation in the frame of density matrix formalism

We have already introduced the density matrix (DM) in (9), and now it is necessary to remind about physical picture behind the elaboration of the DM. First, the DM approach and the Schrödinger equation based one are equivalent and could be transformed one from another [Fano 57], [Fain 72]. A QS interacting with the external fields is considered. The full Hamiltonian is subdivided by three main parts: eigen Hamiltonian $H_0$ which describes internal energy structure of the QS without any external fields plus energy of the electromagnetic field; Hamiltonian of interaction between the QS and the stochastic parts of the external fields $V_{st}$, and Hamiltonian of interaction between the QS and the regular part of the external fields $V_r$. After expansion over the eigen functions of eigen Hamiltonian $H_0$ (particles plus field), the interaction with the stochastic and regular Hamiltonians lead to the qualitatively different terms in the final equations. All interactions with the stochastic fields are considered in the frame of the perturbation theory. The interaction with the vacuum fluctuation ("quantum" part of the stochastic interaction) appears in the first order of perturbation and leads to the relaxation with the time $\tau_{1,r}$ which stands for radiative relaxation time; this is exactly the relaxation time found in (1). Interaction with the classical part of the stochastic field appeared in the second order of the perturbation, because of in the first order the fluctuation with zero mean value (all stochastic parts are assumed to have zero mean value) does not cause any changes – the probability to excite the QS is exactly the same as the probability for the stimulated transition back to ground state. This is an important difference between the vacuum (quantum) field fluctuations and the classical ones, which has been already mentioned above. The transition probability from the ground state with zero photons is zero, while the probability of transition from the excited state with zero photons state is nonzero. In contrast, the probability of transitions with initially nonzero number of photons is nonzero and equal in both directions (for the initial number of photons significantly higher than one), which zeros the effect in the first order of perturbation. The second order, in contrast, gives a nonzero effect, which appears as an extra relaxation time $\tau_{1,nr}$, which gives nonradiative transition rate. The total energy relaxation time is given by:

$$\frac{1}{\tau_1} = \frac{1}{\tau_{1,r}} + \frac{1}{\tau_{1,nr}} \qquad (12)$$

It is important to understand, that the second order of perturbation means self-action, i.e. the QS acts on itself through the in-phase excitation of the external modes. For example, in case of a thermo bath the QS excites some modes which in turn act back on the QS; due to the phase synchronization the action is not zeroed under the averaging.

The approach with the interaction with the thermo bath also allows us to overcome the fundamental problem of losses. In order to get an orthonormal set of the eigen functions (which is widely used), the Hamiltonian has to be hermitian. From the other side, the losses have to be somehow incorporated into the consideration. The model with infinite number of eigen modes, weakly coupled with the considered QS allows to keep the Hamiltonian hermitian, and at the same time introduce the relaxation processes for the QS. In fact, to this extend the density matrix approach is free from the discussed above problem with the imaginary part of permittivity, which makes eigen functions of () nonorthogonal and all consequent conclusions (6) not fully justified.

The second relaxation time $\tau_2$ appears at the consideration of the phase (not amplitude!) changes of the eigen states. In fact, the probability of phase changes without amplitude changes, i.e. without transition between the energy levels, is much higher than the transition between the energy levels. Due to this fact, the phase relaxation time $\tau_2$ appears to be shorter (or equal in the limit) than the energy relaxation time $\tau_1$:

$$\frac{1}{\tau_2} = \frac{1}{\tau_1} + \Delta \qquad (13)$$

It is possible to write down an expression for the extra term $\Delta$ using the respective matrix elements of the stochastic Hamiltonian of interaction [Akulin 87]:

$$\Delta = \frac{1}{\hbar^2} \int w d\Gamma \int_{-\infty}^{\infty} dt \int_{-\infty}^{t} d\tau \left\{ V_{st,1111}(t) V_{st,1111}(\tau) \rho_{11}^{(0)} + V_{st,1112}(t) V_{st,1122}(\tau) \rho_{22}^{(0)} + V_{st,2211}(t) V_{st,2211}(\tau) \rho_{11}^{(0)} + V_{st,2222}(t) V_{st,2222}(\tau) \rho_{22}^{(0)} \right\}$$
(14)

Here $V_{st,ijkl}$ are the matrix elements of the stochastic Hamiltonian of interaction, $\rho_{ij}^{(0)}$ are the unperturbed density matrix elements, $w$ and $d\Gamma$ are the probability of interaction and elementary volume over all parameters of the interaction, which $w$ depends on.

The last Hamiltonian of interaction with the regular fields $V_r$ does not require the perturbation approach and is included in equations (9) rigorously as $H_{12}$. The so called strong coupling regime is one of the phenomena described by this Hamiltonian. From the other side, this interaction can also be treated in the frame of the perturbation approach, but by no means this can be (in general) reduced to the extra relaxation term and packed into the two relaxation times $\tau_1$ and $\tau_2$.

**6. Physical picture of interaction between QS and nanoresonator**

After having all this said, let us consider the physical picture of interaction between the QS and nanoresonator in the frame of the density matrix approach. The presence of the nanoresonator modifies the LDOS, which leads to the modification of the radiative relaxation time ((1), but not (6)). The energy of the excitation of the QS can also be transferred to the stochastic and regular modes of the nanoresonators, coupled to the QS. The stochastic modes are just thermal stochastic electron dynamics, while the regular modes are the collective electron dynamics, i.e. the plasmon modes. In order to subdivide these dynamics (i.e. probability of contribution into stochastic or regular dynamics), the standard approach of probability distribution function (PDF) of the electrons in the nanoresonator $f_{el}(t)$ can be used. The kinetic equation is:

$$\frac{\partial f_{el}}{\partial t} + \left(\vec{v} \cdot \vec{\nabla}_r f_{el}\right) + \frac{q}{m}\left(\vec{E} \cdot \vec{\nabla}_v f_{el}\right) + S = 0 \tag{15}$$

Here $\nabla_r = i\frac{\partial}{\partial x} + j\frac{\partial}{\partial y} + k\frac{\partial}{\partial z}$, $\nabla_v = i\frac{\partial}{\partial v_x} + j\frac{\partial}{\partial v_y} + k\frac{\partial}{\partial v_z}$, $S$ is the impact integral, and $\vec{E}$ is the electric field produced by the QS and is proportional in the near field approximation to the dipole moment of the QS $\vec{E} = \vec{n}_\mu \alpha_{nr} (\rho_{12} + \rho_{21})$, $q$ and $m$ are the charge and mass of an electron respectively. Without an external electric field the *PDF* is just known Maxwell distribution function $f_{el,0} = \left(\frac{m}{2\pi T}\right)^{3/2} \exp\left(-\frac{mv^2}{2T}\right)$, $T$ is the temperature. In case of an external field the *PDF* gets modified with typical response time of $\tau_{el} \sim S^{-1}$, which for typical experimental realizations is about 100 *fs*, i.e. comparable with $\tau_2$. Remind that the introduction of the permittivity and consequently all the concept of the eigen plasmonic functions of the nanoresonator assumes that the *PDF* is already stabilized, i.e. for the times much longer than the 100 *fs*. Taking into account, that the duration of the elementary relaxation process is anyway less or comparable with the 100 *fs*, we come to the conclusion, that the concept of the eigen functions is rigorously speaking not applicable for the description of the relaxation processes. Instead, we might consider a coupled system of equations which includes density matrix and *PDF* equations, namely (9) and (15) and calculate modification of the *PDF* for the respective excited current in the nanoresonator. This excited current influences back the QS, which is actually means second perturbation order or, in other words, self-influence of the QS due to the presence of the nanoresonator. It is also important to include radiative losses into the electron dynamics, i.e. in (15), which can be done by an extra term for the impact integral *S*. This allows us to evaluate the respective contributions to the observed in the tests luminescence enhancement. To the best of our knowledge, this approach to the description of the spontaneous relaxation modification has not yet been realized, but definitively deserves further consideration and discussion.

## 7. On the luminescent enhancement

Now we would like to comment shortly the term of "luminescence enhancement" and discuss the relation between the theoretical picture and observed experiments values. From the point of view of the energy conservation law, any absorbed (pumping) photon can be or reemitted, or relaxed into the nonradiative modes, which corresponds to the final relaxed value of $N = -1$ without pump $W=0$. The radiated intensity dynamics (which is really measured in the tests) in the case of absence of the nanoresonator follows the exponential relaxation law:

$$I_{rad}(t) \sim \frac{(1+N(0))\exp\left(-\frac{t}{\tau_1}\right)-1}{\tau_{1,r}} \tag{16}$$

It is important to point out, that the relaxation dynamics follows the total energy relaxation time $\tau_1$ (not only radiative relaxation time $\tau_{1,r}$!), while an amplitude of the radiated intensity is inversely proportional to the radiative relaxation time $\tau_{1,r}$ only. In case of CW operation mode at $W\tilde{\tau}_1 > 1$, the inversion is $N = N_0$ and the radiated intensity $I_{rad}$ is:

$$I_{rad} \sim \frac{N_0(W)}{\tau_{1,r}} = \frac{W\tilde{\tau}_1 - 1}{\tilde{\tau}_{1,r}} \sim \begin{cases} W, & \tilde{\tau}_{1,r} \ll \tilde{\tau}_{1,nr}, W\tilde{\tau}_{1,r} \gg 1 \\ W\frac{\tilde{\tau}_{1,nr}}{\tilde{\tau}_{1,r}}, & \tilde{\tau}_{1,r} \gg \tilde{\tau}_{1,nr}, W\tilde{\tau}_{1,nr} \gg 1 \end{cases} \tag{17}$$

It is hard to separate experimentally an influence of the pump efficiency and the effects caused by relaxation time changes even in case of the free standing molecules.

## 8. Time dynamics of relaxation in presence of nanoresonator

Let us now consider changes associated with the coupling of the QS and nanoresonator. The system of equations describing the coupled dynamics has been proposed in [Chipouline 12 JOPT] and for the case of the relaxation dynamics is written as:

$$\begin{cases} \dfrac{d\tilde{\rho}_{12}}{dt} + \tilde{\rho}_{12}\left(\dfrac{1}{\tau_2} + i(\omega - \omega_{21})\right) = \dfrac{i\alpha_x \tilde{x}^* N}{\hbar} + \xi_\rho \\ \dfrac{dN}{dt} + \dfrac{(N-N_0)}{\tau_1} = \dfrac{i\alpha_x\left(\tilde{x}\tilde{\rho}_{12} - \tilde{x}^*\tilde{\rho}_{12}^*\right)}{2\hbar} \\ 2(\gamma - i\omega)\dfrac{d\tilde{x}}{dt} + \left(\omega_0^2 - \omega^2 - 2i\omega\gamma\right)\tilde{x} = \alpha_\rho \tilde{\rho}_{12}^* + \xi_x \end{cases} \tag{18}$$

Here both relaxation times $\tau_1$ and $\tau_2$ are assumed to be modified due to the presence of the extra stochastic interaction due to the presence of the nanoresonator:

$$\begin{cases} \dfrac{1}{\tau_1} = \dfrac{P_r}{\tau_{1,r}} + \dfrac{P_{nr}}{\tau_{1,nr}} \\ \dfrac{1}{\tau_2} = \dfrac{1}{\tau_1} + \Delta(P) \end{cases} \tag{19}$$

According to (14), the extra part differentiating energy and phase relaxation time $\Delta(P)$ is also modified, which is underlined here by the dependence on $P$ ($P$ - Purcell effect). Method of the calculations or estimations for the both Purcell factors $P_r$ and $P_{nr}$ is not discussed here, as well as the estimation of $\Delta(P)$. An exact solution for the relaxation dynamics of (18) cannot be found analytically, and is supposed in general to be calculated numerically. The results of the typical dynamics are presented in Fig. 1.

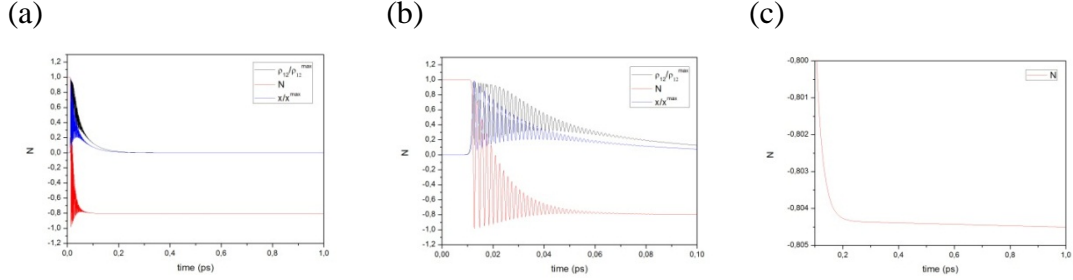

**Fig. 1: Relaxation dynamics of the system of coupled QS and plasmonic nanoresonator at full resonance $\omega = \omega_0 = \omega_{12}$. The dynamics (a) consists of fast oscillation at the times less than the phase relaxation time (b) and the exponential relaxation for the rest time (c). Please, keep in mind the different scales in (a), (b), and (c).**

The dynamics is clearly subdivided by two stages: the first one for the time less than the phase relaxation time $\tau_2$ exhibits Rabi oscillations, typical for the strong coupling, and the rest is just trivial exponential relaxation dynamics. This behaviour has been recently experimentally measured in [Vasa 13]. Note that the density matrix approach allows us to consider strong and weak coupling in the frame of the same model, namely in the frame of (18): these two regimes correspond to high and low values of the coupling constants $\alpha_\rho$ and $\alpha_x$; of course, each set of the coupling coefficients corresponds to the respective set of Purcell factors $P_r$ and $P_{nr}$. Both the coupling coefficients and Purcell factor values depend on the distance and mutual positions of the QS and nanoresonator; the method of estimations for these coefficients (for example, used in [Stockman 11]) is out of the scopes of this paper and will be discussed elsewhere.

The dynamics of (18) is calculated for the initial conditions $N(t=0) > 0$ (inversion) and $N_0 = -1$ (absence of pump). For the case of the CW operation mode it is assumed that $N_0 > 0$ (permanent pump) and $N$ (and other two variables $\rho_{12}$ and $x$) are assumed to be found for the stationary case, i.e. under assumption $\frac{d}{dt} = 0$. As it has been discussed in previous chapters, this system of equations describes the stationary and stochastic (with extra Langevin terms) dynamics of the nanolaser/spaser [Bergman 06], [Stockman 10], [Stockman 11]. It has to be realized, that system (18) possesses two types of solutions – regular and stochastic. The regular nontrivial solution exists only for the pump rates above the threshold one $W > W_{th}$; otherwise the regular solution is zero. In contrast, random fluctuations, described by correlators $\langle x(\omega) x^*(\omega') \rangle$, $\langle \rho_{12}(\omega) \rho_{12}^*(\omega') \rangle$ and $\langle x(\omega) \rho_{12}^*(\omega') \rangle, \langle x^*(\omega) \rho_{12}(\omega') \rangle$ are nonzero and give extra factors to the estimation for the luminescent enhancement. These calculations are also left for the future publications; here it is worth noting that the problem of spontaneous emission enhancement in the frame of the developed approach is methodologically equivalent to the estimation of the spontaneous emission from any laser below generation threshold, and to this extend is not new.

Qualitatively, the luminescent intensity modification can be caused by several factors. First, at the resonance between nanoresonator eigen mode and the pump wavelengths any nanoresonator works as an optical antenna [Novotny 11], which enhances pump rate efficiency $W$ and therefore increases $N_0(W)$ in (17); that coupling between the QS and nanoresonator out of resonance is small and could be neglected to the first approximation. But even in the case of resonance with the emission wavelength (and thus nonresonant for pumping wavelength), the effective pump rate is also enhanced by the presence of the metallic nanoresonator. According to (17), both nominator and denominator turn out to be modified and it is extremely hard to separate experimentally these two effects. The enhancement from the coupling with the regular parts of the electron dynamics is proportional to the correlations $\langle x(\omega)\rho_{12}^*(\omega')\rangle, \langle x^*(\omega)\rho_{12}(\omega')\rangle$ (see second equation in (18)) and could be roughly estimated by the relation of the phase and the total relaxation times $\tau_2/\tau_1$ which is about $10^{-3}$. It means that the part of the energy, emitted due to the resonant interaction between the QS and nanoresonator is very small in compare with the total emitted energy and could be safely neglected. The only case when this interaction could place a significant role is the nanoresonator enhanced Dike effect, when all energy due to synchronization between the QS will be emitted in a one short pulse with the duration comparable with the phase relaxation time. Probability of the Dike effect in the considered coupled system is very low at the typical experimental conditions and the pumping rate below the generation threshold. From the other side, above threshold the QS are synchronized due to the nanoresonator (as it takes place in typical "macro" laser), and this state corresponds to the stationary solution of (18) (see also [Dorofeenko 13] about Dike effect for the arrays of the coupled nanolasers/spasers). Neglecting the coherent contribution to the luminescent intensity modification, the only factor (excepting already mentioned pump efficiency modification) is the two Purcell factors $P_r$ and $P_{nr}$, causing radiative and nonradiative relaxation time modification (19) which are supposed to be substituted in (17) in order to estimate the effect.

## 9. Conclusion

In conclusion of this paper, I would like to summarize the discussion in following statements:

1. I have considered qualitatively the commonly accepted method of estimation of the spontaneous relaxation dynamics. While the method is undoubtedly justified for the spontaneous relaxation in a free space, application of the same procedure for the case of the nanoresonators optically coupled to the QS is questionable. The concerns are:
   a) Different orders of magnitude for the perturbation expansion in the model, which describe principally different physical phenomena, namely coupling with the vacuum fluctuations and self-action at the coupled dynamics with the free electrons in the nanoresonator.
   b) An application of the eigen mode formalism causes a question about necessary time required for the mode formation; otherwise the interaction with stochastic modes of the free electrons appears to be more qualitatively appropriate.

c) An application of the eigen mode formalism fails for the highly loss materials, i.e. for the metallic nanoresonators due to the nonhermitian nature of the helmholz operator and consequently nonorthogonal eigen functions.
   d) In the frame of the accepted in textbook [Novotny 06] approach, there is no subdivision between interaction with stochastic and regular fields, which definitively leads to the different relaxation dynamics.
2. The density matrix approach has been analyzed in compare with the commonly accepted one. It has been shown, that this approach is qualitatively more adequate and free from the mentioned above concerns. It has been suggested to include in consideration one more dynamical equation for the electron distribution function. It is expected, that this approach allows us to estimate the introduced here Purcell factors $P_r$ and $P_{nr}$ for both radiative and nonradiative relaxation times.
3. The proposed system of equations allows us to calculate regular and stochastic dynamics above and below the lasing threshold, and consider both strong and weak coupling regimes in the frame of the same approach.

REFERENCES

1. [Purcell 46] E. Purcell, Phys. Rev., 69, 681, 1946.
2. [Klimov 01] V. Klimov, M. Ducloy, V. Letokhov, Quantum Electronics 31(7), 596, 2001.
3. [Giannini 11] V. Giannini, A. Fernandez-Dominquez, S. Heck, S. Maier, Chem. Phys., 111, 3888, 2011.
4. [Dregely 12] D. Dregely, K. Lindfors, J. Dorfmüller, M. Hentschel, M. Becker, J. Wratchup, M. Lippitz, R. Vogelgesang, H. Gissen, Phys. Status Solidi B 249, 666, 2012.
5. [Berman 94] P. Berman (Ed.) Cavity Quantum Electrodynamics (New York: Academic, 1994)
6. [Haroche 92] S. Haroche, in Fundamental System in Quantum Optics. J.Dalibard, J.M.Raimond, J.Zinn-Justin (Eds) (Les Houches, 1990; Elsevier Science Publishers B V, 1992)
7. [Belacel 13] C. Belacel, B. Habert, F. Bigourdan, F. Marquier, J.-P. Hugonin, S. Michaelis de Vasconcellos, X. Lafosse, L. Coolen, C. Schwob, C. Javaux, B. Dubertret, J.-J. Greffet, P. Senellart, and A. Maitre, Nano Lett., 13, 1516, 2013.
8. [Esteban 10] R. Esteban, T.V. Teperik, and J. J. Greffet, PRL 104, 026802, 2010.
9. [Vasa 13] P. Vasa, W. Wang, R. Pomraenke, M. Lammers, M. Maiuri, C. Manzoni, G. Cerullo and C. Lienau, NATURE PHOTONICS, 7, 128, 2013.
10. [Vlack 12] C. Van Vlack, Philip Trøst Kristensen, and S. Hughes, Phys. Rev B **85**, 075303, 2012.
11. [Trügler 08] A. Trügler and U. Hohenester, Phys. Rev. B **77**, 115403, 2008.
12. [Filter 12] R. Filter, S. Mühlig, T. Eichelkraut, C. Rockstuhl, and F. Lederer, Phys. Rev B **86**, 035404, 2012.
13. [Kern 12_1] A. M. Kern and O. J. F. Martin, Phys. Rev A **85**, 022501, 2012.
14. [Stobbe 12] S. Stobbe, P. T. Kristensen, J. E. Mortensen, J. M. Hvam, J. Mork, and P. Lodahl, Phys. Rev. B **86**, 085304, 2012.
15. [Kern 11] A. M. Kern and O. J. F. Martin, Nano Lett., 11, 482, 2011.

16. [Kern 12_2] A. M. Kern, Alfred J. Meixner, and O. J. F. Martin, ASCNano, 6, 11, 9828, 2012.
17. [Caze 12] A. Cazé, R. Pierrat, R. Carminati, Photonics and Nanostructures – Fundamentals and Applications, 10, 339, 2012.
18. [Kort 13] W. J. M. Kort-Kamp, F. S. S. Rosa, F. A. Pinheiro, and C. Farina, Phys. Rev. A **87**, 023837, 2013.
19. [Poddubny 12] A. Poddubny, P. Belov, P. Ginzburg, A. Zayats, and Y. Kivshar, Phys. Rev. B **86**, 035148, 2012.
20. [Poddubny 13] A. Poddubny, P. Belov, and Y. Kivshar, Phys. Rev. B **87**, 035136, 2013.
21. [Barthes 11] J. Barthes, G. Colas des Francs, A. Bouhelier, J.-C. Weeber, and A. Dereux, Phys. Rev. B **84**, 073403, 2011.
22. [Maragkou 12] M. Maragkou, A. K. Nowak, E. Gallardo, H. P. van der Meulen, I. Prieto, L. J. Martinez, P. A. Postigo, and J. M. Calleja, Phys. Rev. B **86**, 085316, 2012.
23. [Ruppin 82] R. Ruppin, J. Chem. Phys. 76(4), 1982.
24. [Gersten 81] J. Gersten, A. Nitzan, J. Chem. Phys. 75(3), 1981.
25. [Koenderink 10] A. F. Koenderink, Opt. Lett. 35, 4208, 2010.
26. [Mertens 07] H. Mertens, A. F. Koenderink, and A. Polman, Phys. Rev. B **76**, 115123, 2007.
27. [Ginzburg 83], V. Ginzburg, UFN 140, 535, 1983 [in Russian].
28. [Weisskopf 35] Naturwissenschaften, 27, 631, 1935.
29. [Weisskopf 81] Physics Today, v. 34, No. 11, p. 69, 1981.
30. [Fermi 32] Rev. Mod. Phys., 4, 87, 1932.
31. [Novotny 06] L. Novotny, B. Hecht, Principles of Nano-Optics; Cambridge University Press: New York, 2006.
32. [Dung 00] H. Dung, L. Knoll, D-G. Welsch, Phys. Rev. A 62, 3804, 2000.
33. [Knoll 00] L. Knoll, S. Scheel, D-G. Welsch, Quant-ph/000621-27 June, 2000.
34. [Yeung 96] M. S. Yeung and T. K. Gustafson, Phys. Rev. A, 54, 6, 1996.
35. [Fano 57] U. Fano, "Description of States in Quantum Mechanics by Density Matrix and Operator Techniques". In: Rev. Mod. Phys. 29, 74, 1957.
36. [Fain 72] V. Fain, "Quantum radiophysics: photons and nonlinear media", Sovetskoe Radio, 1972 [in Russian].
37. [Akulin 87] V. Akulin, N. Karlov, "Intensive resonance interaction in quantum electronics", Nauka, 1987 [in Russian].
38. [Chipouline 12 JOPT] A. Chipouline, S. Sugavanam, V. A. Fedotov, A. E. Nikolaenko, "Analytical model for active metamaterials with quantum ingredients", J. Opt. **14** 114005, 2012.
39. [Novotny 11] L. Novotny and N. van Hulst, Nature Photonics, 5, 83, 2011.
40. [Dorofeenko 13] A. Dorofeenko, A. Zyablovsky, A. Vinogradov, E. Andrianov, A. Pukhov, and A. Lisyansky, OPTICS EXPRESS 21, 14539, 2013.
41. [Stockman 11] M. Stockman, "Spaser action, loss-compensation, and stability in plasmonic systems with gain", PRL **106**, 156802, 2011.
42. [Bergman 03] D. J. Bergman and M. I. Stockman, PRL **90**, 027402, 2003.
43. [Stockman 10] M. I. Stockman, "The Spaser as a Nanoscale Quantum Generator and Ultrafast Amplifier", Journal of Optics 12, 024004-1-13, 2010.